\documentstyle[floats,aps,eqsecnum]{revtex}

\begin{document}
\draft

\catcode`\@=11 \catcode`\@=12 
\twocolumn[\hsize\textwidth\columnwidth\hsize\csname@twocolumnfalse\endcsname
\title{Effective mass of composite fermion: a phenomenological fit
in with anomalous propagation of surface acoustic wave }

\author{Yue Yu and Zhaobin Su}

\address{Institute of Theoretical Physics, Chinese Academy of Sciences,
 Beijing 100080, P. R. China}

\maketitle
\begin{abstract}

We calculate the conductivity associated with the anomalous propagation 
of a surface acoustic wave above a two-dimensional electron gas 
at $\nu=1/2$. Murthy-Shankar's middle representation is adopted 
and a contribution to the response functions beyond the random phase approximation 
is taken into account. We give a phenomenological 
fit for the effective mass of composite fermion in with the 
experimental data of the anomalous propagation of surface acoustic wave
at $\nu=1/2$ and find the phenomenological value of the effective mass is 
several times larger than the theoretical value 
$m_{th}^*=6\varepsilon/e^2l_{1/2}$ derived from the Hartree-Fock 
approximation. We compare the phenomenological value of 
the composite fermion effective mass with that measured in the experiments of
the activation energy and the Shubnikov-de Haas oscillations. It is found that 
the comparison is fairly well.

\end{abstract}

]

\section{Introduction}

The composite fermion (CF) theory for fractional quantum Hall effect
has been evolved for a decade since 
Jain first proposed this complex of the electron and flux quanta in
connection with the trial wave functions of the prominent quantum Hall
states \cite{Jain}. An important step in the development of the CF theory was
stridden by Haplerin, Lee and Read (HLR) \cite{HLR} as well as Kalmeyer and
Zhang \cite{KZ}, in which the fermion-Chern-Simons (FCS) theory \cite{FL} is
applied to describe the physics at the even-denominator filling fractions,
e.g., $\nu= \frac{1}{2}$. HLR successfully predicted the existence
of the CF Fermi surface at such filling fractions, which indicated a CF
Fermi liquid or a modified Fermi liquid. However, HLR recognized a crucial
difference between the FCS theory and the conventional Fermi liquid, namely the
effective mass of CF is divergent at the Fermi surface. This draws forth many
subsequent studies \cite{book}. Most recently, it was recognized that the charge carried by the CF is obviously different from the real physical excitation at
$\nu=1/2$ \cite{read}. This stirs up a
series of recent studies \cite{read,rs,wy,ms,ms1,ysd,lee,hald,ssh}. Of them, the
dipolar neutral CF is attracting particular attentions from many authors,
which stems from Read's conceptual paper \cite{read}. Based on the neutral
CF, the effective mass problem was carefully reconsidered \cite{ms,hald,ysd,ms2}%
. Shankar and Murthy found a way to define the CF effective mass and showed that it
is of the order of the electron-electron interaction, namely, $1/m^*\simeq
(1/\nu)^{2/3}e^2 l_B/6\varepsilon$, in the Hartree-Fock approximation \cite{ms}.

Beside the prediction of the CF Fermi surface, another celebrated success of
HLR's theory is the explanation for the anomaly of the surface acoustic
wave (SAW) propagation above a two-dimensional electron gas in GaAs/AlGaAs
heterostructures \cite{will}. In their experiment, Willett et al observed the anomalous maximum and minimum at $\nu=\frac{1}{2}$ of the
attenuation rate and the velocity shift of the SAW , respectively \cite{will2}.
Correspondingly, a maximum of the conductivity appears at $\nu=\frac{1}{2}$. 
In the clear
range where the wavelength of the SAW is shorter than the CF's mean free
path, the conductivity has a linear-dependence on the wave vector. The theoretical
results from the FCS theory agree qualitatively with these experimental
observations \cite{HLR}. However, from the beginning, there is a
systematic discrepancy between the theoretical and experimental values for
the magnitude of the conductivity $\sigma_{xx}(q)$. In the FCS theory, there
is no adjustable parameter to enhance $\sigma_{xx}(q)$ such that its theoretical
value is approximately a factor of 2 smaller than the
conductivity observed in experiments \cite{HLR}. We notice that this
non-adjustability in the FCS theory is due to the perturbative random phase
approximation (RPA). For a generic field theory, the RPA may or may not 
provide a full description for the
low energy behavior of the system . Particularly, for the FCS theory, the important
low energy information might be lost by only taking the RPA. It is because,
firstly, the Chern-Simons coupling constant, which is one of the parameters in
the perturbative expansion, is not small; Secondly, the fluctuations of the
Chern-Simons gauge field is subject to the well-known constraint for the
physical states which will be shown explicitly below in (\ref{cst}). Murthy and
Shankar have recently given their RPA calculation according to their `final
representation' and found a good fit in with the SAW experiment \cite{ms}.
This is a tempting result, but there are also some dangers, e.g., it will result
in the incompressibility at $\nu=1/2$. Although those dangerous points could be
clarified \cite{sh,ssh}, the restoration of the compressibility does not help to
solve the above mentioned discrepancy. Therefore, in the present paper, 
we would like to go beyond the RPA and consider further a non-RPA interaction 
which is ignored in Murthy-Shankar's
discussions. We calculate a bubble diagram of the Chern-Simons gauge fluctuation
(Fig.1) and its contribution to the response function.  We find that this 
contribution is comparable to that from the free CF response function 
in their magnitude but with opposite signs. As a result,  $
\sigma_{xx}(q)$ is enhanced by an adjustable parameter which comes from Fig.1.
This parameter is related to the CF effective mass. Hence,
using the experimental data in the SAW propagation \cite
{will,will2,will3,will4,will5}, one can fit the effective mass and finds
that it is in several times of Shankar-Murthy's theoretical
estimation, $m^*=(e^2l_{1/2}/6\varepsilon)^{-1}$. 
We argue that it is reasonable that the effective mass
increases for $\nu\to 1/2$ because,
theoretically, the gauge fluctuations always raise the effective mass with respect
to its Hartree-Fock value and experimentally, the increase of the effective
mass has been observed much faster than the current theoretical predictions
\cite{du,du1,lead}. We also compare our phenomenological effective mass
 to those in measurements of the activation
energy and the Shubnikov-de Haas oscillation \cite{du2,man,du,du1,lead}.

This paper is organized as follows. In Sec. II, we review 
Murthy-Shankar's theory and set up Feynman's rules for the
perturbative theory. In Sec. III, various response functions are calculated.
In Sec. IV, the conductivity including a non-RPA correction is derived. In
Sec. V, the CF effective mass is fit in with the SAW propagation experiments
and is compared to the known theoretical and experimental results. The
section VI consists of our conclusions.

\section{Hamiltonian and Perturbative Theory}

There are essentially two kinds of formulations in the study of
the CF theory. One of them is proposed, by construction,
within the subspace of the lowest Landau level (LLL) 
\cite{read,hald,lee}. These formulations
automatically incorporate the feature that the energy scale and effective
mass are set by the electron-electron interaction. The other kind of
formulations begins with an enlarged quantum state space. The FCS theory is the
basis of the later one. In the present paper, we shall relate the
effective mass to the conductivity in the SAW propagation. The FCS theory
already gave the anomaly of the SAW propagation a qualitative description.
We, thus, prefer to use a formulation starting from a variant of the FCS Hamiltonian,
which is proposed by Murthy and Shankar \cite{ms,ms1}. Enlightened by Bohm and Pines
\cite{bp},  Murthy and Shankar chose a gauge for 
the FCS Hamiltonian  and named this Hamiltonian the `middle representation' of their theory. 
For the CF excitation with a correct charge, they applied
a canonical transformation to the middle representation and arrived at their
`final representation'. In their final representation, the CF excitation is
neutral at $\nu=1/2$ and the effective mass is of the order of the
electron-electron interaction and independent of the band mass. On the other
hand, it has been shown that if we do not set foot in the quasiparticle
charge, we can also get this same value of the effective mass by using the middle
representation \cite{ysd}. In this paper, we would like to fit the effective
mass in with the experiment of the SAW propagation. Thus, we will employ
Murthy-Shankar's middle representation.

\subsection{Hamiltonian}

We start from a two dimensional interacting electron system that is placed
in a uniform magnetic field $B$ perpendicular to the two dimensional plane
imbedded in a uniform positive background. We assume that all electrons are
spin-polarized. For the two-body interaction potential $V$, the $N$-electron
Hamiltonian reads,

\begin{equation}
H_e=\frac{1}{2m_b}\sum_i\biggl[-i\hbar\nabla_i+\frac{e}{c}\vec{A}_i(\vec{x}%
_i) \biggr]^2+\sum_{i<j} V(\vec{x}_i-\vec{x}_j),
\end{equation}
where the vector potential $\vec{A}$ corresponds to the magnetic field $B$
and $m_b$ is the band mass of the electrons. Hereafter, we will use the unit 
$\frac{e}{c}=\hbar=1$. At this stage, we do not confine the electrons in the
LLL. The attraction between the electrons and the uniform background is not
explicitly shown up.

Following a common treatment, we make an anyon transformation \cite{lmw} for
the electron wavefunction $\Phi (\vec r_1,...,\vec r_N)$ with $\vec r_j$ being the
position of the $j$-th electron. The transformed wavefunction is given by
\begin{equation}
\Psi _{cs}(z_1,...,z_N)=\prod_{i<j}\biggl[\frac{z_i-z_j}{|z_i-z_j|}\biggl]^{%
\tilde \phi }\Phi (z_1,...,z_N),
\end{equation}
where $z_j=x_j+iy_j$, and $\tilde \phi $ is an even integer. $\Psi _{cs}$ is the
wavefunction for the transformed fermion (so called the Chern-Simons fermion). The Hamiltonian corresponding to the transformation becomes 
\begin{equation}
H_{cs}=\frac{1}{2m_b}\sum_i\biggl[-i\nabla _i+\vec A_i(\vec x_i)-\vec a_i( 
\vec x_i)\biggr]^2+\sum_{i<j}V(\vec x_i-\vec x_j),  \label{csh}
\end{equation}
where $\vec a$ is a statistical gauge potential. i.e., 
\begin{equation}
\vec a(\vec x_i)=\frac{\tilde \phi }{2\pi }\sum_{j\not{=}i}\frac{\hat z%
\times (\vec x_i-\vec x_j)}{|\vec x_i-\vec x_j|^2},  \label{sg}
\end{equation}
which satisfies the following constraint 
\begin{equation}
\nabla \times \vec a(\vec x)=2\pi \tilde \phi \rho (\vec x)\equiv b(\vec x).
\label{cst}
\end{equation}

The mean field approximation for the FCS theory can be
achieved as 
\begin{equation}
\vec a_{MF}=\vec A,
\end{equation}
where $\vec A=(B/2)\hat z\times \vec x$. Around the mean field state, there
is an important gauge fluctuation $\delta \vec a=\vec A-\vec a$. 
For convenience, we denote $\delta \vec a$ as $\vec a$ which is the
fluctuation of the statistic gauge field around the mean field. If we introduce
the second quantization notation with the Chern-Simons fermion field $\psi
_{cs}$, then the Hamiltonian around the mean field reads, 
\begin{eqnarray}
H&=&\int d^2x\frac 1{2m_b}\biggl|(-i\nabla +\vec a(\vec x))\psi _{cs}\biggr| %
^2  \nonumber \\
&+& \frac 12\int d^2xd^2x^{\prime }\delta \rho (\vec x)V(\vec x-\vec x
^{\prime })\delta \rho (\vec x^{\prime }).
\end{eqnarray}
with $\delta \rho =\rho -\rho _0$. The gauge fluctuation obeys the
constraint (\ref{cst}) associated with the density fluctuation $\delta \rho$ .
Notice that here the Hamiltonian is written in the Coulomb gauge $\nabla
\cdot \vec a=0$. It is well-known that the FCS theory has a gauge symmetry
corresponding to the gauge transformation of $\vec a$ if we consider the
bulk states only so that the
Halmitonian can also be written as a gauge invariant form. Then, one can choose other gauges to deal with the system.
For our purpose, we choose a gauge that was used by Shankar and Murthy 
\cite{ms}, enlightened by Bohm-Pines' gauge choice for the three dimensional
electron gas in the real electromagnetic field \cite{bp}. In this gauge, 
the Hamiltonian, what is called the middle representation, takes its form as
\begin{equation}
H_{cf}=H_{0f}+H_{0a}+H_{i}+H_{ia}+H_{sr},
\end{equation}
where $H_{sr}$ is the non-dynamic short-range gauge fluctuation and 
\begin{eqnarray}
H_{0f}&=&\frac{1}{2m_b}\int d^2 x |\nabla\psi|^2,  \nonumber \\
H_{0a}&=&\frac{\rho_0}{2m_b}\int d^2 x(a_x^2+a_y^2)  \nonumber \\
&+&\frac{1}{8\pi^2 \tilde\phi^2}\int d^2xd^2x^{\prime}[\nabla\times\vec{a}(%
\vec{x})]V(\vec{x}-\vec{x}^{\prime})[\nabla^{\prime}\times\vec{a} (\vec{x}%
^{\prime})],  \nonumber \\
H_{i}~&=&\int d^2 x \vec{a}\cdot \frac{i}{2m_b}(\psi^\dagger\nabla
\psi-\nabla\psi^\dagger \psi),  \nonumber \\ 
H_{ia}&=&\frac{1}{2m_b}\int d^2x \delta\rho\vec{ a}^2.  \label{hdc}
\end{eqnarray}
Here $H_{0f}$ and $H_{0a}$ stand for the free Hamiltonian of the CF and the
gauge fluctuation, respectively. $H_i$ and $H_{ia}$ are the interactions.
The Fourier component $\vec{a}$ of the gauge field obeys the commutation relation 
\begin{equation}
[a_x(\vec{q}), a_y(\vec{q}^{\prime}]=i\delta^{(2)}(\vec{q}-\vec{q}^{\prime}),
\end{equation}
and is restricted by the constraint 
\begin{equation}
(\frac{qa_x}{2\pi l_{1/2}}-\delta \rho)|{\rm Phys}>=0,~~0<q<Q.
\end{equation}
with the cut-off $Q=k_F$ \cite{ms}. All short range fluctuations are 
included in $H_{sr}$, which is ignored in discussing the low energy physics we
are interested in. Murthy-Shankar's formulation begins
with the above Hamiltonian after dropping $H_{sr}$. Finally, they
ignored the non-RPA interaction $H_{ia}$ and employed a canonical
transformation to cancel $H_i$ and reach their final representation. Here,
instead of using the final representation, we exert their middle
representation which is favorite for the perturbative calculation. However,
to directly deal with $H_{ia}$ in the perturbative theory is difficult.
Solving $\delta \rho$ by the constraint (\ref{cst}), $H_{ia}$ is transformed 
into 
\begin{equation}
H^{\prime}_{ia}=\frac{1}{4\pi\tilde\phi m_b}\int d^2x (\nabla\times \vec{a}) 
\vec{a}^2.
\end{equation}
Now, we have the Hamiltonian 
\begin{equation}
H=H_{0f}+H_{0a}+H_i+H^{\prime}_{ia}
\end{equation}
to work on.

\subsection{ Feynman's Rules}

The perturbative theory starts from to set up Feynman's rules. Since we have
neglected the short wave length gauge fluctuations, the gauge field wave
vector is restricted to $q<k_F$ in all following perturbative calculation.
The free CF propagator (Fig. 2(a)) is 
\begin{equation}
G_0(k,\omega )=\frac{\theta (k-k_F)}{\omega -\epsilon _k+i0^{+}}+\frac{%
\theta (k_F-k)}{\omega -\epsilon _k-i0^{+}},
\end{equation}
and the gauge fluctuation propagates (Fig. 2(b)) can be read out  
\begin{equation}
D^0(q,\omega )=U,
\end{equation}
with the matrix $U$ defined by 
\begin{equation}
U^{-1}=\left( 
\begin{array}{cc}
\displaystyle -\frac{\rho _0}{m^*} & {\displaystyle \frac{-i\omega }{2\pi 
\tilde \phi }} \\ 
{\displaystyle\frac{i\omega }{2\pi \tilde \phi }} & \displaystyle -\frac{%
\rho _0}{m^*}(1+\frac{e^2q}{2\omega _c\varepsilon }) \\ 
& 
\end{array}
\right) .  \label{GP}
\end{equation}
Here we have taken the $2\times 2$ matrix description of the gauge
propagator with $D^0_{11}=U_{11}$ and $D^0_{22 }=U_{22}$ and so on. The
interaction has been specified as the Coulomb interaction $V(q)=\frac{2\pi e^2%
}{\varepsilon q}$, which is taken throughout in the whole paper. The indices
`1' and `2'correspond to the components parallel and perpendicular to the wave
 vector $\vec{q}$, respectively. The interaction vertex is shown as (Fig. 2(c)) 
\begin{equation}
g_a=\frac 1{m^*}((\vec k+\frac{\vec q}2)\cdot \hat q,(\vec k+\frac{\vec q}2%
)\times \hat q).\label{ga}
\end{equation}
In eqs. (\ref{GP}) and (\ref{ga}), we have simply replaced the band mass by a phenomenological effective mass as did by HLR \cite{HLR}.
The gauge field self-interaction vertex figured in Fig. 2(d) can be read out from the Hamiltonian
\begin{eqnarray}
&&f_{122}(-\vec{q}-\vec{q},\vec{q}, \vec{q}^{\prime}) =\frac{i}{8\pi m_b}%
(q_2+q_2^{\prime}),  \nonumber \\
&&f_{211}(-\vec{q}-\vec{q},\vec{q}, \vec{q}^{\prime}) =\frac{-i}{8\pi m_b}%
(q_1+q_1^{\prime}).  \label{NRPA}
\end{eqnarray}
We call the interaction described by the vertex (\ref{NRPA}) the non-RPA
interaction. It has been pointed out that this interaction vertex is not
renormalizable. So, neither the band mass in (\ref{NRPA}) because it is not
relevant to the CF kinetic energy \cite{ms}.

\section{response functions}

\subsection{Non-interacting Response Functions}

The calculation of the response functions is the central task of this
section. The simplest response functions are  the free CF's (Fig. 3), which
are defined as \cite{HLR} 
\begin{eqnarray}
K_{00}^0&=&\int \frac{d^2k}{(2\pi)^2}\frac{f(\omega_{\vec k+\vec q/2}
-f(\omega_{\vec k-\vec q/2})} {\omega-\omega_{\vec k+\vec q/2}+\omega_{\vec k%
-\vec q/2}+i\delta}, \\
K_{11}^0&=&\int \frac{d^2k}{(2\pi)^2}|\frac{k_\parallel}{m^*}|^2 \frac{%
f(\omega_{\vec k+\vec q/2}-f(\omega_{\vec k-\vec q/2})} {\omega-\omega_{\vec %
k+\vec q/2}+\omega_{\vec k-\vec q/2}+i\delta},  \nonumber \\
K_{22}^0&=&\int \frac{d^2k}{(2\pi)^2}|\frac{k_\perp}{m^*}|^2 \frac{f(\omega_{%
\vec k+\vec q/2}-f(\omega_{\vec k-\vec q/2})} {\omega-\omega_{\vec k+\vec q%
/2}+\omega_{\vec k-\vec q/2}+i\delta},  \nonumber
\end{eqnarray}
where $f(\omega_k)$ is the Fermi factor. In the current-current response
function, we do not subtract $\frac{\rho_0}{m^*}$ because it has appeared in
(\ref{GP}). The static response functions for $q<<k_F$ are calculated as 
\begin{eqnarray}
K_{00}^0(q,0)&=&\frac{m^*}{2\pi}+O(q^2), \label{tt} \\
K_{11}^0(q,0)&=&\frac{\rho_0}{m^*}+O(q^4),  \nonumber \\
K_{22}^0(q,0)&=&\frac{\rho_0}{m^*}-\frac{q^2}{24\pi m^*}+O(q^4).  \nonumber
\end{eqnarray}
For $\omega<<v_Fq$, the imaginary parts of $K^0$ are calculated as follows: 
\begin{eqnarray}
{\rm Im}K_{00}^0(q,\omega)&\approx& \frac{m^*}{2\pi}\frac{\omega}{v_F q},
\label{im} \\
{\rm Im}K_{11}^0(q,\omega)&\approx& \frac{m^*}{2\pi}\frac{\omega}{v_F q} 
\frac{\omega^2}{q^2},  \nonumber \\
{\rm Im}K_{22}^0(q,\omega)&\approx&\frac{2\rho_0\omega}{k_Fq}.  \nonumber
\end{eqnarray}
Meanwhile there is a non-zero real part of $K_{11}^0$ 
\begin{equation}
{\rm Re}K_{11}^0(q,\omega)\approx \frac{m^*}{2\pi}\frac{\omega^2}{q^2}.
\label{re}
\end{equation}
It is seen that 
\begin{equation}
K^0_{11}-\frac{\rho_0}{m^*}=\frac{\omega^2}{q^2}K^0_{00},
\end{equation}
which simply recovers the physics of the continuous equation $j_1=(\omega/q)j_0$. Comparing the imaginary ones with the real ones in eqs.(\ref{tt})-(\ref{re}),
${\rm Im}K_{00}^0$ and ${\rm Im}K_{11}^0$ are
small and can be neglected while ${\rm Im}K_{22}^0$ should be kept. 

On the other hand, for $\omega>>v_Fq$, it is easy to have 
\begin{equation}
K_{11}^0\sim K_{22}^0\sim O(\frac{(v_Fq)^2}{\omega^2}).
\end{equation}

\subsection{Bare Non-RPA Response Functions}

Another kind of the response functions which may be relevant is 
so-called bare non-RPA response functions (see Fig. 4). 
These response functions are determined
by the non-RPA interaction vertex (\ref{NRPA}) and the bare gauge
fluctuation propagator $D_0$ (see Fig. 2(b)). The calculating results for
these response functions in $q\ll k_F$ are as follows. 
Because there is no pole for $\omega\ll v_F q$ in the integration, one can easily
see that
\begin{equation}
K_{b,11}^0(q,0)=K_{b,22}^0(q,0)=K_{b,12}^0(q,0)=0,
\end{equation}
i.e., these static response functions  vanish. For $\omega>>v_F q$, one has 
\begin{eqnarray}
K_{b,11}^0(q,\omega)&=&K_{b,22}^0(q,\omega)=-\frac{q^2}{16\pi m_b} \frac{1}{%
4-z^2},  \nonumber \\
K_{b,12}^0(q,\omega)&=&i\frac{q^2}{32\pi m_b}\frac{z}{4-z^2},
\end{eqnarray}
with $z=\omega/\omega_c$. Because the poles of the bare non-RPA response
functions are in the high energy region, they will not contribute to the low energy
behavior.

\subsection{RPA Response Functions and RPA Gauge Propagator}

The RPA equation for $K$ (Fig. 5) may be written as 
\begin{equation}
K_R=K^0-K^0[K^0+U^{-1}]^{-1}K^0.
\end{equation}
According to the calculations in the previous two subsections, we see that
the low frequency limit of $K_R$ is not modified if one replaces $K^0$ by $%
K^0+K^0_b$. Meanwhile, although there is the anomalous pole $\omega=2\omega_c
$ in the bare non-RPA response functions for $\omega>>v_Fq$, they will not
violate Kohn's theorem since they are as small as the order of $q^2$.

Therefore, the bare non-RPA response functions will not affect the physical
properties we are interested in. According to such a result , the RPA gauge propagator $D^r$ (the thick wave line in Figures) can
also be defined by neglecting $K^0_b$. That is,  $D^r=(K^0+U^{-1})^{-1}$. For the $\omega<<v_Fq$ case it can be written as 
\begin{eqnarray}
&&D^r_{11}(\vec{q},\omega)=\frac{2\pi}{m^*}\frac{q^2}{(\omega+i\delta)^2}, 
\nonumber \\
&&D^r_{22}(\vec{q},\omega)=\frac{q}{i\omega\gamma-q^2\chi},  \label{prop} \\
&&D^r_{12}(\vec q,\omega)=-D^r_{12}=\frac{iq^2}{2m^*}\frac{1}{\omega+i\delta}
\frac{1}{i\omega\gamma-q^2\chi},
\end{eqnarray}
where $\gamma=\frac{k_F}{2\pi}$ and $\chi=\frac{e^2}{8\pi\varepsilon}$.

\subsection{Response Functions beyond RPA}

So far, we find no interesting result beyond the
RPA although the bare non-RPA response functions are calculated in subsection B.
We did not find out a nontrivial contribution to the response
functions for $\omega<<v_Fq$. On the other hand, we calculated the RPA gauge
propagator in the last subsection. One may ask what could be happened 
if the bare gauge propagator in the bubble of the bare non-RPA response 
function is replaced by the RPA
gauge propagator?  Instead of Fig. 4, a response function beyond the RPA is
represented as Fig. 1 and defined by 
\begin{eqnarray}
K^{nr}_{aa^{\prime}}(q,\omega)&=&\int \frac{d^2q^{\prime}}{(2\pi)^2}\frac{%
d\omega^{\prime}}{2\pi i} f_{abc}f_{a^{\prime}b^{\prime}c^{\prime}} 
\nonumber \\
&\times&D^r_{bb^{\prime}}(\vec q+\vec q^{\prime},\omega+\omega^{%
\prime})D^r_{cc^{\prime}}(\vec q^{\prime},\omega^{\prime}).
\end{eqnarray}
Before going to the details, we recall some useful properties: 
The gauge propagator $D^r_{12}=-D^r_{21}$ and the coupling constants
$f_{112}(q^\prime,\theta)f_{122}(q^\prime,\theta)=
-f_{112}(q^\prime,-\theta)f_{122}(q^\prime,-\theta)$. Those properties
immediately lead to $%
K^{nr}_{12}=K^{nr}_{21}=0$. So, we are only interested in $K^{nr}_{aa}$ for $a=1,2
$, e.g., 
\begin{eqnarray}
K^{nr}_{22}&=&\int \frac{d^2 q^{\prime}}{(2\pi)^2}\frac{d\omega^{\prime}}{%
2\pi i} \\
&&[-(f_{211}(\vec q^{\prime}+\vec q))^2D^r_{11}(\vec q+\vec q%
^{\prime},\omega+\omega^{\prime}) D^r_{11}(\vec q^{\prime},\omega^{\prime}) 
\nonumber \\
&&- (f_{212}(\vec q^{\prime}+\vec q))^2D^r_{11}(\vec q+\vec q%
^{\prime},\omega+\omega^{\prime}) D^r_{22}(\vec q^{\prime},\omega^{\prime}) 
\nonumber \\
&&-(f_{221}(\vec q^{\prime}+\vec q))^2D^r_{22} (\vec q+\vec q%
^{\prime},\omega+\omega^{\prime})D^r_{11}(\vec q^{\prime},\omega^{\prime})],
\nonumber
\end{eqnarray}
where we have dropped many vanishing terms (e.g., the terms with respect to
$D^r_{12}$ due to the property $D^r_{12}=-D^r_{21}$, etc).
Substituting the non-RPA
interaction vertex and the RPA gauge propagator into the above equation, and 
after a variable shift, the leading term in the case $q<<k_F$ can be
derived as
\begin{eqnarray}
K^{nr}_{22}&\approx&\frac{2\pi}{m^*}\frac{1}{(8\pi m_b)^2}\int \frac{d^2
q^{\prime}}{(2\pi)^2}\frac{d\omega^{\prime}}{2\pi i} \\
&&{q^{\prime}_2}^2{q^{\prime}}^3\biggl[\frac{1}{(\omega^{\prime}+\omega+i%
\delta)^2} +\frac{1}{(\omega^{\prime}-\omega-i\delta)^2}\biggr]  \nonumber \\
&&\times \frac{1}{i\gamma\omega^{\prime}-\chi {q^{\prime}}^2}-O(\frac{%
\omega^2}{k_F^2}),  \nonumber
\end{eqnarray}
where the factor $q^{'2}_2$ comes from $f_{221}^2=f_{212}^2$ and $q^{'3}$ from the
propogators; the $q$-dependent contributions are of the higher order.
After the integration, one can attract a pure imaginary 
constant contribution to $K^{nr}_{22}$, 
\begin{equation}
K^{nr}_{22}\approx -\frac{ik_F Q^3\varepsilon^2}{12\pi m_b^2m^* e^4} (1-O(%
\frac{\omega^2}{k_F^2})),  \label{knr}
\end{equation}
where $Q$ is the cut-off in integrating over $q^{\prime}$ and is taken to
be $Q=k_F$ according to the gauge choice. $O(\frac{\omega^2}{k_F^2})$ is
positive so that it reduces a small value of $K^{nr}_{22}$. Because (\ref{knr})
is invariant if one replaces $q_2^{\prime 2}$ by $q_1^{\prime 2}$, a similar
calculation shows that
\begin{equation}
K^{nr}_{11}=K^{nr}_{22},
\end{equation}
in their leading terms. Our calculation provides a correction to the imaginary
part of the response function. In some sense, we may use 
\begin{equation}
K^m=K^0+K^{nr},  \label{KM}
\end{equation}
to replace $K^0$.

\section{The conductivity tensor}

The conductivity tensor is one of the most important observables in the
experiment of the transport property of the system. For the FCS system, the
conductivity tensor has been defined by Halperin, Lee and Read in their
seminal paper \cite{HLR}. We here take their definition with a minor
modification. Basically, HLR's definition of the conductivity tensor is
valid for $\omega\sim v_s q$ with $v_s$ being the SAW propagation velocity and
the tensor reads 
\begin{eqnarray}
\frac{1}{\sigma_{xx}(\vec q, \omega)}&=&\frac{iq^2}{\omega}\biggl [\frac{1}{%
\Pi_{00}(q,\omega)}-\frac{1}{\Pi_{00}(q,0)}\biggr],  \nonumber \\
\sigma_{yy}(\vec q, \omega)&=&-\frac{i}{\omega}[\Pi_{22}(\vec q,\omega ) -%
{\rm Re}(\Pi_{22}(\vec q,0))],  \label{cpi} \\
\sigma_{xy}(\vec q, \omega)&=&-\sigma_{yx}(\vec q, \omega)= \frac{i}{q}%
\Pi_{02}(\vec q,\omega),  \nonumber
\end{eqnarray}
where a modification has been made in defining $\sigma_{yy}$ with the real
part of $\Pi_{22}(\vec q,0)$ being subtracted. Similarly, the ``quasiparticle
conductivity tensor" is defined by 
\begin{eqnarray}
\frac{1}{\tilde\sigma_{xx}(\vec q, \omega)}&=&\frac{iq^2}{\omega}\biggl [%
\frac{1}{\tilde K_{00}(q,\omega)}-\frac{1}{\tilde K_{00}(q,0)}\biggr], 
\nonumber \\
\tilde\sigma_{yy}(\vec q, \omega)&=&-\frac{i}{\omega}[\tilde K_{22}(\vec q%
,\omega ) -{\rm Re}(\tilde K_{22}(\vec q,0))],  \label{ctk} \\
\sigma_{xy}(\vec q, \omega)&=&-\sigma_{yx}(\vec q, \omega)= \frac{i}{q}%
\tilde K_{02}(\vec q,\omega),  \nonumber
\end{eqnarray}
The matrix $\Pi$ in (\ref{cpi}) consists of the sum of all Feynman diagrams
for the full response function matrix $K$ which is irreducible with respect
to the Coulomb interaction, while $\tilde K$ in (\ref{ctk}) contains only
those diagrams which are irreducible with respect to both the Chern-Simons
interaction and the Coulomb interaction. In the regime where the
definition of the conductivity tensors are available, $\sigma$ and $\tilde
\sigma$ are related to the corresponding resistivities $\rho$ and $\tilde\rho$ through the matrix equations 
\begin{eqnarray}
&&\sigma\equiv \rho^{-1},  \nonumber \\
&&\rho=\tilde\rho+\rho_{cs},  \nonumber \\
&&\tilde\rho\equiv\tilde\sigma^{-1}, \\
&&\rho_{cs}=\frac{4\pi\hbar}{e^2}\left( 
\begin{array}{cc}
0 & -1 \\ 
1 & 0
\end{array}
\right),  \nonumber
\end{eqnarray}
respectively.
Considering the case where $\tilde\sigma_{xy}=- \tilde\sigma_{yx}=0$, we
have 
\begin{eqnarray}
&&\tilde\rho_{xx}(\vec q,\omega)=\frac{1}{\tilde \sigma_{xx}(\vec q, \omega)}%
,  \nonumber \\
&&\tilde\rho_{yy}(\vec q,\omega)=\frac{1}{\tilde \sigma_{yy}(\vec q, \omega)}%
.
\end{eqnarray}

Using the above definition of the conductivity tensor, the formula for $%
\sigma_{xx}(q)$ may then be rewritten as 
\begin{equation}
\sigma_{xx}(q)={\rho_{yy}(q)}/{\rho_{xy}^2},
\end{equation}
where $\rho_{xy}=4\pi\hbar/e^2$, and 
\begin{equation}
\frac{1}{\rho_{yy}(q)}=e^2\lim_{\omega=v_sq\to 0} \frac{1}{\omega} {\rm Im} 
\tilde K_{22}(q,\omega).  \label{rhoyy}
\end{equation}
It is impossible to get the exact form of $\tilde K$. The RPA definition of the 
conductivity tensor consists of replacing
$\tilde K$ by the free CF response function $K^0$. We may
improve the RPA definition by replacing $\tilde K$ by $K^m$. Thus, we have, 
\begin{equation}
\frac{1}{\rho_{yy}(q)}=e^2\lim_{\omega=v_sq\to 0} \frac{1}{\omega} {\rm Im}
K^m_{22}(q,\omega).  \label{rhoyym}
\end{equation}

\section{Surface Acoustic Wave Propagation}

\subsection{Anomaly of the Surface Acoustic Wave Propagation}

The experimental observation for the SAW propagation exhibits an anomaly in
the relevant conductivity which deviates from the normal macroscopic DC value
\cite{will}. This phenomenon may be observed through the measurement of the SAW velocity
shift $\Delta v_s$ and the attenuation rate $\kappa$ for the SAW amplitude, 
\begin{eqnarray}
&&\frac{\Delta v_s}{v_s}=\frac{\alpha^2}{2}\frac{1}{1+[\sigma_{xx}(q)/
\sigma_m]^2},  \nonumber \\
&&\kappa=\frac{q\alpha^2}{2}\frac{[\sigma_{xx}(q)/\sigma_m]}{
1+[\sigma_{xx}(q)/\sigma_m]^2},
\end{eqnarray}
where $\sigma_m=\frac{v_s \varepsilon}{2\pi}$ and $\alpha$ is a constant
proportional to the piezoelectric coupling of GaAs. The experiments of the
SAW's propagation were performed in high quality GaAs/AlGaAs
heterostructures with $q\ll k_F$ and $\omega=v_s q\ll v_Fq$. The experiment
result shows that the longitudinal conductivity is linearly dependent on the
SAW's wavevector \cite{will}, which is qualitatively in agreement with the prediction
of the FCS theory approach \cite{HLR}. However, to compare in details with the
experimental data, it is necessary to use a value of $\sigma_m$
approximately four times larger than the theoretical one. This discrepancy
remains not explained so far. Moreover, there is an additional quantitative
discrepancy between the theoretical prediction and experimental data in the
longitudinal conductivity. In the RPA, HLR arrived at \cite{HLR} 
\begin{eqnarray}
\rho_{yy}(q)&=&\frac{2\pi}{k_F}q\frac{\hbar}{e^2}, ~~{\rm for}~~ q\gg\frac{2%
}{l},  \label{clear} \\
\rho_{yy}(q)&=&\frac{4\pi}{k_Fl}\frac{\hbar}{e^2}, ~~{\rm for}~~ q\ll \frac{2%
}{l},  \nonumber
\end{eqnarray}
where $l$ is the CF transport mean free path at $\nu= \frac{1}{2}$. As 
emphasized by the authors of ref.\cite{HLR} 
there is no adjustable parameters in the RPA of (\ref{rhoyy} ) while the
theoretical values of $\sigma_{xx}$ are approximately a factor of 2 smaller
than the experimental values obtained by Willett et al. That is, the
theoretical result (\ref{clear}) is `too small' in a factor of 2. We notice that
the unadjustability comes from the replacement of $\tilde K$ by $K^0$ in the
RPA. To make it be adjustable, we suggest that 
$\tilde K$ in (\ref{rhoyy}) should be approximated by $
K^m$ instead of $K^0$. Hence, the inverse of the transverse resistivity of
the CF in the clear range is given by 
\begin{eqnarray}
\frac{1}{\rho_{yy}(q)}&=&e^2\lim_{\omega=v_sq\to 0} \frac{1}{\omega} {\rm Im}%
K^m_{22}(q,\omega)  \nonumber \\
&\approx& \frac{k_F}{4\pi q}(2-C)\frac{e^2}{\hbar},~~~{\rm for}
~~q>>\frac{2}{l}, \label{nrhoyy}
\end{eqnarray}
where the physical dimension has been restored and the constant $C$ is defined by
\begin{equation}
C=\frac{v^*_F/v_s}{3(m_b/m_{coul})^2}= \frac{k_F\hbar/m^*v_s}{%
3(m_b/m_{coul})^2}  \label{C}
\end{equation}
with $m_{coul}=\frac{\varepsilon k_F\hbar^2}{e^2}$ being a mass scale induced by
the Coulomb interaction. Different from the result given by (\ref{clear}),
therefore, (\ref{nrhoyy}) includes an adjustable parameter $C$ in the $q$%
-dependent conductivity $\sigma_{xx}(q)$. When the experimental parameters
are fixed, $C $ is entirely determined by the CF effective mass. To be fit in with
these series of experimental results, we should have 
$C\approx 1$. Then, we have to check with the CF effective mass to see whether it is consistent with such a fit or not.

\subsection{Phenomenological Fit of the Effective Mass}

In this subsection, we use the experiment data in a set of the SAW
propagation experiments to fit the CF effective mass and compare the fit
result to the established values of $m^*$ both theoretically and
experimentally.

For GaAs/AlGaAs heterostructures, we take the dielectric constant $%
\varepsilon=12.6$ and the electron band mass $m_b\approx 0.07 m_e$. We refer
to several sets of experimental data by Willett et al as follows.

\noindent (A) The two-dimensional electron gas density $n_e=6.6\times 10^{10}{\rm cm}%
^{-2}$, the frequency of SAW $f=2.4{\rm GHz}$ and the corresponding
wavelength $\lambda =1.2\mu {\rm m}$ \cite{will};

\noindent (B) $n_e=6\times 10^{10} {\rm cm}^{-2}$, $f=1.5 {\rm GHz}$ and $\lambda=2.0
\mu {\rm m}$ \cite{will2};

\noindent (C) $n_e=7\times 10^{10} {\rm cm} ^{-2} $, $f=0.36 {\rm GHz}$ and $%
\lambda=7.8\mu {\rm m}$ \cite{will5};

\noindent (D) $n_e=1.0\times 10^{11} {\rm cm}^{-2}$, $f=1.2 {\rm GHz}$ and $%
\lambda=8\mu {\rm m}$ \cite{will3};

\noindent (E) $n_e=1.6\times 10^{11} {\rm cm}^{-2}$, $f=10.7 {\rm GHz}$ and $%
\lambda=0.27 \mu {\rm m}$ \cite{will4}.

First, let us compare $m^*$ with the theoretical effective mass $m^*_{th}$.
We take the theoretical value of the CF effective mass to be the 
Hartree-Fock one \cite{ms,ysd}, i.e., 
\begin{equation}
m^*_{th}=6 m_{coul},
\end{equation}
which is in good agreement with the result of the numerical simulation \cite
{MdA}. If one takes $C\approx 1$ in (\ref{C}), one gets the effective mass
corresponding to the experiment data, 
\begin{eqnarray}
&&m^*_{(A)}\approx 2.50m^*_{th},~~m^*_{(B)}\approx 2.17m^*_{th},~~
m^*_{(C)}\approx 2.67m^*_{th},  \nonumber \\
&&m^*_{(D)}\approx 3.00m^*_{th},~~m^*_{(E)}\approx 6.00m^*_{th},  \label{tth}
\end{eqnarray}
where $m^*_{(A)}$ corresponds to the experimental data in (A) etc.

The experimental data of the effective mass from Willett et al are not
available. To have an instructive understanding, we compare the
phenomenological effective mass with the effective mass measured in other
experiments. In the activation energy gap measurement, Du et al \cite{du}
gave $m^*_{exp}=0.57 m_e$ and Manoharran et al gave $m^*_{exp}=1.4 m_e$ \cite
{man}. (Here $m_e$ is the bare mass of the electron.) The Shubnikov-de Haas
effective provides $m^*_{exp}=0.7 m_e$ \cite{du1} and even a much larger
value (divergence) \cite{du2} by Du et al while Leadley et al gave $%
m^*_{exp}=0.51 m_e$ \cite{lead}. Our phenomenological results are 
\begin{eqnarray}
&&m^*_{(A)}=0.91m_e, ~~m^*_{(B)}=0.74 m_e,~~ m^*_{(C)}=0.99 m_e,  \nonumber
\\
&& m^*_{(D)}=1.34 m_e,~~m^*_{(E)}=3.38 m_e, \label{texp}
\end{eqnarray}
Examining (\ref{tth}), the phenomenological fit of the effective mass seems
to be `heavier' in several times than theoretical one. There may be several
possible explanations to this result:

(1). It is possible that there are still some important diagrams that are
not taken into account to approximate $K$ and are expected to further improve our
calculation.

(2). The higher order terms are neglected in the calculation of $K^0$ and $%
K^{nr}$, which may give a further adjustment for the effective mass. For example, $%
O(\omega^2/k_F^2)$ in (\ref{knr}) decreases $K^{nr}$ such that $m^*$
decreases in order to keep $C=1$.

(3). The larger effective mass at $\nu=1/2$ coincides with the increase
(even divergence) of the effective mass at this filling fraction.

(4). The theoretical value of the effective mass should be, in fact, larger
than Hartree-Fock one because the gauge fluctuations always enlarge the
effective mass. The experimental measurement of the effective mass supports
this point because all $m^*_{exp}$ we quoted are larger than $m^*_{th}$ in a
ratio $m^*_{exp}/m_{th}^*\geq 1.58$.

On the other hand, the comparison between the phenomenological effective
mass showed in (\ref{texp}) and the experimental values is fairly well
although $m^*$ is little bit larger than $m^*_{exp}$. However, $m_{(E)}^*$
seems to be extraordinarily large. The reason for this remains unknown.

\section{conclusions}

In conclusions, we have shown a possibility to resolve the discrepancy
between theory and experiment in the longitudinal conductivity dependent on
the wavevector of the SAW propagating above the 2DEG. The key point is to
take a non-RPA correction to the response function from the self-interaction
among the gauge fluctuations into account. This correction increases the
theoretical value of the longitudinal conductivity and gives an adjustable
parameter which relates the CF effective mass to the conductivity. Thus, the
phenomenological effective mass of CF is estimated by using the data in the
SAW propagation experiments and one found that it is fairly consistent with
the effective mass measured in the experiments relating to the activation
energy and the Shubnikov-de Haas oscillations. Although the phenomenological
value of the CF effective mass is several times larger than the theoretical
one obtained by a Hartree-Fock approximation, we explained the possible sources to
cause this.

\bigskip

\noindent{\large \bf Acknowledgments}

\bigskip

The authors would like to thank Dr. Xi Dai for the useful discussions. 
This work was supported in part by the NSF of China.



\centerline{\bf FIGURE CAPTIONS}

Fig. 1: One-loop diagram of fluctuations from the self interaction of the
gauge field, where the gauge propagator is taken to be the RPA one.

Fig. 2: (a) The free CF propagator; (b) The bare gauge propagator; (c) The
CF-gauge fluctuation interaction vertex; (d) The self-interaction vertex of
the gauge fluctuations.

Fig. 3: The non-interaction CF response function.

Fig. 4: The bare non-RPA response function.

Fig. 5: The RPA response function.

\end{document}